\documentclass[11pt]{article}
\usepackage{times}
\usepackage{geometry}
\usepackage[utf8]{inputenc}
\usepackage{enumitem,amssymb}
\usepackage{ragged2e}
\usepackage{graphicx}
\usepackage{comment}
\usepackage{multicol}
\usepackage[usenames]{xcolor} 
\usepackage{wrapfig}

\definecolor{xlinkcolor}{cmyk}{1,1,0,0}
\usepackage{url}
\usepackage[
 colorlinks=true,    
 linkcolor=xlinkcolor,     
 citecolor=xlinkcolor,     
 filecolor=xlinkcolor,  
 urlcolor=xlinkcolor,      
 final=true
]{hyperref}
\usepackage[super,sort&compress]{natbib}
\usepackage{enumitem}
\setenumerate{itemsep=0mm}

\begin{document}
\begin{raggedright} 
\huge
Snowmass2021 - Letter of Interest \hfill \\[+1em]
\textit{Cosmology Intertwined III: $f \sigma_8$ and $S_8$} \hfill \\[+1em]
\end{raggedright}

\normalsize

\noindent {\large \bf Thematic Areas:}  (check all that apply $\square$/$\blacksquare$)

\noindent $\blacksquare$ (CF1) Dark Matter: Particle Like \\
\noindent $\square$ (CF2) Dark Matter: Wavelike  \\ 
\noindent $\square$ (CF3) Dark Matter: Cosmic Probes  \\
\noindent $\blacksquare$ (CF4) Dark Energy and Cosmic Acceleration: The Modern Universe \\
\noindent $\square$ (CF5) Dark Energy and Cosmic Acceleration: Cosmic Dawn and Before \\
\noindent $\square$ (CF6) Dark Energy and Cosmic Acceleration: Complementarity of Probes and New Facilities \\
\noindent $\blacksquare$ (CF7) Cosmic Probes of Fundamental Physics \\
\noindent $\square$ (Other) {\it [Please specify frontier/topical group]} \\

\noindent {\large \bf Contact Information:}\\
Eleonora Di Valentino (JBCA, University of Manchester, UK) [eleonora.divalentino@manchester.ac.uk]\\

\noindent {\large \bf Authors:}  \\[+1em]
Eleonora Di Valentino (JBCA, University of Manchester, UK)\\
Luis A. Anchordoqui (City University of New York, USA)\\
\"{O}zg\"{u}r Akarsu (Istanbul Technical University, Istanbul, Turkey) \\
Yacine Ali-Haimoud (New York University, USA)\\
Luca Amendola (University of Heidelberg, Germany)\\
Nikki Arendse (DARK, Niels Bohr Institute, Denmark) \\
Marika Asgari (University of Edinburgh, UK)\\
Mario Ballardini (Alma Mater Studiorum Universit\`a di Bologna, Italy)\\
Spyros Basilakos (Academy of Athens and Nat. Observatory of Athens, Greece) \\
Elia Battistelli (Sapienza Universit\`a di Roma and INFN sezione di Roma, Italy)\\
Micol Benetti (Universit\`a degli Studi di Napoli Federico II and INFN sezione di Napoli, Italy)\\
Simon Birrer (Stanford University, USA)\\
Fran\c{c}ois R. Bouchet (Institut d'Astrophysique de Paris, CNRS \& Sorbonne University, France) \\
Marco Bruni (Institute of Cosmology and Gravitation, Portsmouth, UK, and INFN Sezione di Trieste, Italy)\\
Erminia Calabrese (Cardiff University, UK)\\
David Camarena (Federal University of Espirito Santo, Brazil) \\
Salvatore Capozziello (Universit\`a degli Studi di Napoli Federico II, Napoli, Italy) \\
Angela Chen (University of Michigan, Ann Arbor, USA)\\
Jens Chluba (JBCA, University of Manchester, UK)\\
Anton Chudaykin (Institute for Nuclear Research, Russia) \\
Eoin \'O Colg\'ain (Asia Pacific Center for Theoretical Physics, Korea) \\
Francis-Yan Cyr-Racine (University of New Mexico, USA) \\
Paolo de Bernardis (Sapienza Universit\`a di Roma and INFN sezione di Roma, Italy) \\
Javier de Cruz P\'erez (Departament FQA and ICCUB, Universitat de Barcelona, Spain)\\
Jacques Delabrouille (CNRS/IN2P3, Laboratoire APC, France \& CEA/IRFU, France \& USTC, China)\\
Jo Dunkley (Princeton University, USA)\\
Celia Escamilla-Rivera (ICN, Universidad Nacional Aut\'onoma de M\'exico, Mexico) \\
Agn\`es Fert\'e (JPL, Caltech, Pasadena, USA)\\
Fabio Finelli (INAF OAS Bologna and INFN Sezione di Bologna, Italy) \\
Wendy Freedman (University of Chicago, Chicago IL, USA)\\
Noemi Frusciante (Instituto de Astrof\'isica e Ci\^encias do Espa\c{c}o, Lisboa, Portugal)\\
Elena Giusarma (Michigan Technological University, USA) \\
Adri\`a G\'omez-Valent (University of Heidelberg, Germany)\\
Will Handley (University of Cambridge, UK) \\
Ian Harrison (JBCA, University of Manchester, UK) \\
Luke Hart (JBCA, University of Manchester, UK)\\
Alan Heavens (ICIC, Imperial College London, UK)\\
Hendrik Hildebrandt (Ruhr-University Bochum, Germany)\\
Daniel Holz (University of Chicago, Chicago IL, USA)\\
Dragan Huterer (University of Michigan, Ann Arbor, USA)\\
Mikhail M. Ivanov (New York University, USA) \\
Shahab Joudaki (University of Oxford, UK and University of Waterloo, Canada) \\
Marc Kamionkowski (Johns Hopkins University, Baltimore, MD, USA) \\
Tanvi Karwal (University of Pennsylvania, Philadelphia, USA) \\
Lloyd Knox (UC Davis, Davis CA, USA)\\
Suresh Kumar (BITS Pilani, Pilani Campus, India) \\
Luca Lamagna (Sapienza Universit\`a di Roma and INFN sezione di Roma, Italy) \\
Julien Lesgourgues (RWTH Aachen University) \\
Matteo Lucca (Universit\'e Libre de Bruxelles, Belgium)\\
Valerio Marra (Federal University of Espirito Santo, Brazil) \\
Silvia Masi (Sapienza Universit\`a di Roma and INFN sezione di Roma, Italy) \\
Sabino Matarrese (University of Padova and INFN Sezione di Padova, Italy) \\
Arindam Mazumdar (Centre for Theoretical Studies, IIT Kharagpur, India) \\
Alessandro Melchiorri (Sapienza Universit\`a di Roma and INFN sezione di Roma, Italy)\\
Olga Mena (IFIC, CSIC-UV, Spain)\\
Laura Mersini-Houghton (University of North Carolina at Chapel Hill, USA) \\
Vivian Miranda (University of Arizona, USA) \\
Cristian Moreno-Pulido (Departament FQA and ICCUB, Universitat de Barcelona, Spain)\\
David F. Mota (University of Oslo, Norway) \\
Jessica Muir (KIPAC, Stanford University, USA)\\
Ankan Mukherjee (Jamia Millia Islamia Central University, India) \\
Florian Niedermann (CP3-Origins, University of Southern Denmark) \\
Alessio Notari (ICCUB, Universitat de Barcelona, Spain) \\
Rafael C. Nunes (National Institute for Space Research, Brazil)\\
Francesco Pace (JBCA, University of Manchester, UK)\\
Andronikos Paliathanasis (DUT, South Africa and UACh, Chile) \\
Antonella Palmese (Fermi National Accelerator Laboratory, USA) \\
Supriya Pan (Presidency University, Kolkata, India)\\
Daniela Paoletti (INAF OAS Bologna and INFN Sezione di Bologna, Italy)\\
Valeria Pettorino (AIM, CEA, CNRS, Universit\'e Paris-Saclay, Universit\'e de Paris, France) \\
Francesco Piacentini (Sapienza Universit\`a di Roma and INFN sezione di Roma, Italy)\\
Vivian Poulin (LUPM, CNRS \& University of Montpellier, France) \\
Marco Raveri (University of Pennsylvania, Philadelphia, USA) \\
Adam G. Riess (Johns Hopkins University, Baltimore, USA) \\
Vincenzo Salzano (University of Szczecin, Poland)\\
Emmanuel N. Saridakis (National Observatory of Athens, Greece)\\
Anjan A. Sen (Jamia Millia Islamia Central University New Delhi, India) \\
Arman Shafieloo (Korea Astronomy and Space Science Institute (KASI), Korea)\\
Anowar J. Shajib (University of California, Los Angeles, USA) \\
Joseph Silk (IAP Sorbonne University \& CNRS, France, and Johns Hopkins University, USA)\\
Alessandra Silvestri (Leiden University, NL)\\
Martin S. Sloth (CP3-Origins, University of Southern Denmark) \\
Tristan L. Smith (Swarthmore College, Swarthmore, USA)\\ 
Joan Sol\`a Peracaula (Departament FQA and ICCUB, Universitat de Barcelona, Spain)\\
Carsten van de Bruck (University of Sheffield, UK) \\
Licia Verde (ICREA, Universidad de Barcelona, Spain)\\
Luca Visinelli (GRAPPA, University of Amsterdam, NL) \\
Benjamin D. Wandelt (IAP Sorbonne University \& CNRS, France, and CCA, USA) \\
Deng Wang (National Astronomical Observatories, CAS, China) \\
Jian-Min Wang (Key Laboratory for Particle Astrophysics, IHEP of the CAS, Beijing, China) \\
Anil K. Yadav (United College of Engg. \& Research, GN, India)\\
Weiqiang Yang (Liaoning Normal University, Dalian, China) \\

\noindent {\large \bf Abstract:} 
The standard $\Lambda$ Cold Dark Matter cosmological model provides a wonderful fit to current cosmological data, but a few tensions and anomalies became statistically significant with the latest data analyses. While these anomalies could be due to the presence of systematic errors in the experiments, they could also indicate the need for new physics beyond the standard model. In this Letter of Interest we focus on the tension of the Planck data with weak lensing measurements and redshift surveys, about the value of the matter energy density $\Omega_m$, and the amplitude or rate of the growth of structure ($\sigma_8,f\sigma_8$). We list a few interesting models for solving this tension, and we discuss the importance of trying to fit with a single model a full array of data and not just one parameter at a time.

\begin{wrapfigure}{R}{0.45\textwidth}
\centering
\includegraphics[width=0.4\textwidth]{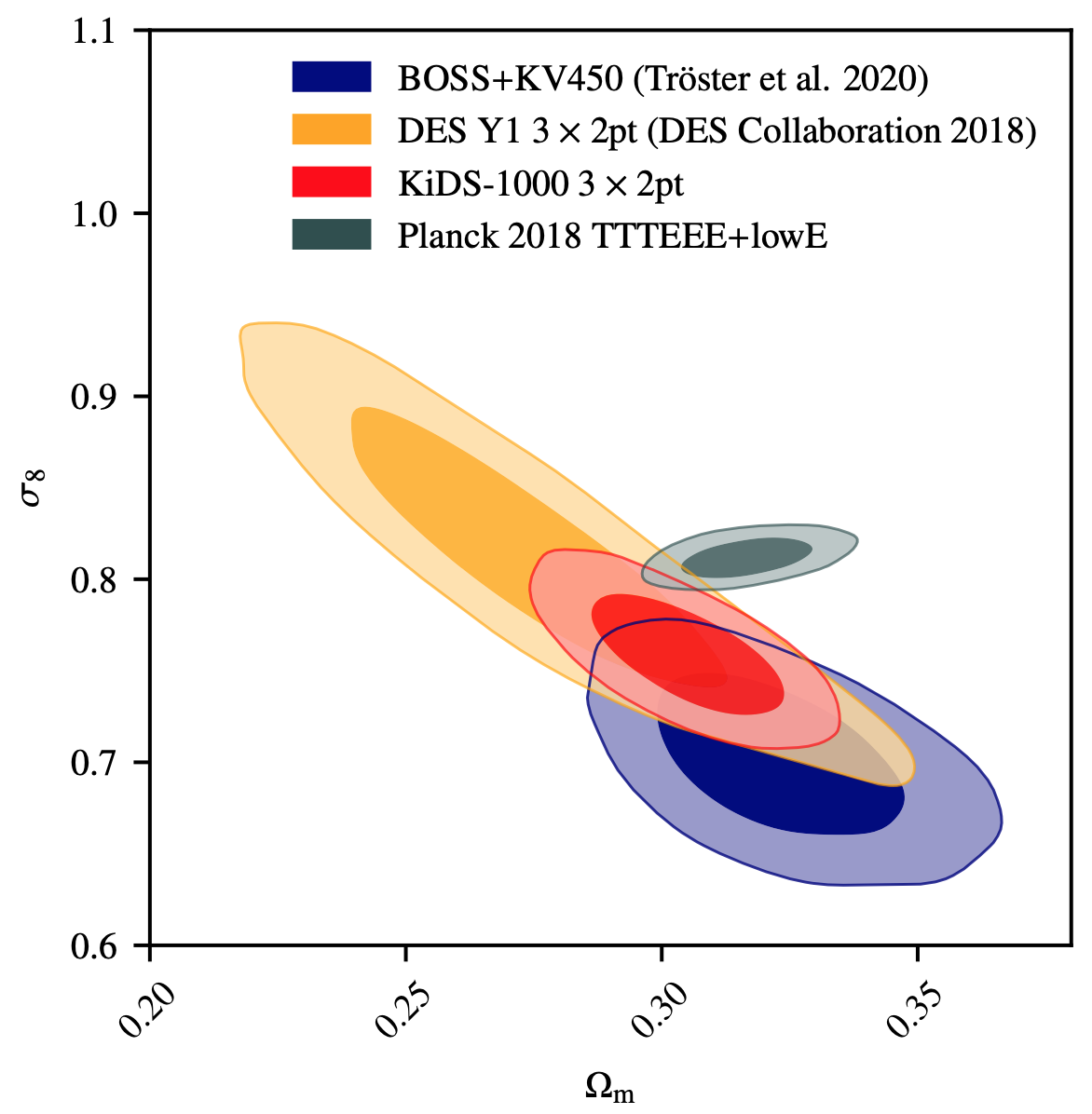}
\caption{68\% CL and 95\% CL contour plots for $\sigma_8$ and $\Omega_m$(from Ref.~\cite{Heymans:2020gsg}).}
\label{2D}
\end{wrapfigure}

\clearpage
\noindent {\bf The $S_8$ tension --} 
The standard $\Lambda$ Cold Dark Matter ($\Lambda$CDM) cosmological model provides an amazing fit to current cosmological data. However, some statistically-significant tensions in cosmological parameter estimations emerged between the Planck experiment, measuring the Cosmic Microwave Background (CMB) anisotropies, and other low-redshift cosmological probes. In addition to the long standing {\it Hubble constant} $H_0$ disagreement, a tension of the Planck data with weak lensing measurements and redshift surveys has been reported, about the value of the matter energy density $\Omega_m$, and the amplitude or rate of growth of structure ($\sigma_8,f\sigma_8$). Although this tension could be due to systematic errors, it is worthwhile to investigate the possibility of new physics beyond the standard model. The tension can be visualized in the $\sigma_8$ vs $\Omega_m$ plane (see Fig.~\ref{2D}) and is often quantified using the $S_8 \equiv \sigma_8 \sqrt{\Omega_{m}/0.3}$ parameter, along the main degeneracy direction of weak lensing measurements. This can be also related to $f\sigma_8(z=0)$, measured by galaxy redshift space distortions (RSD)~\cite{Li:2016bis,Gil-Marin:2016wya}, where $f=[\Omega_m(z)]^{0.55}$ approximates the growth rate.

\begin{wrapfigure}{R}{0.45\textwidth}
\centering
\includegraphics[width=0.4\textwidth]{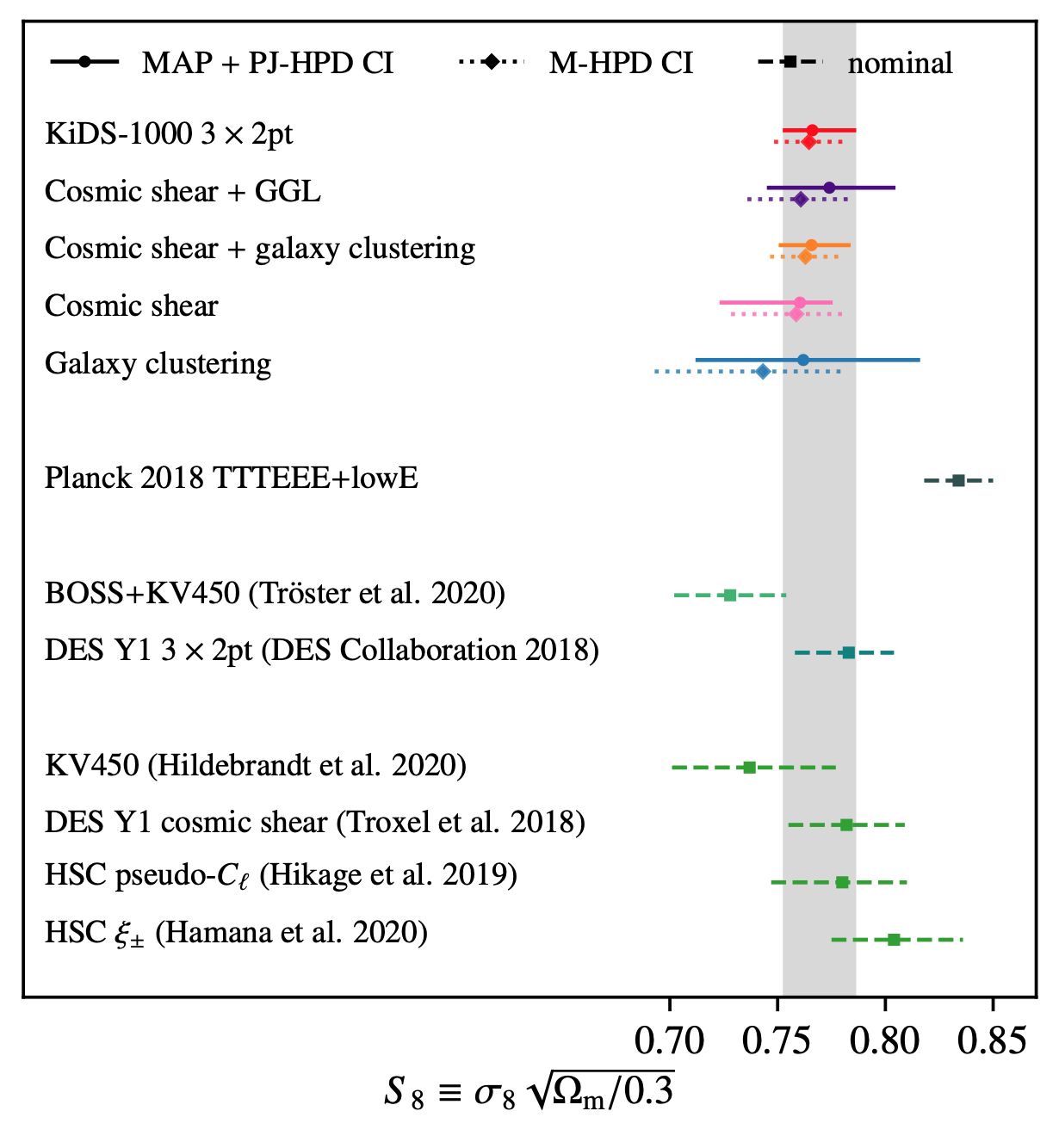}
\caption{Whisker plot showing the 68\% error bars on $S_8$ (from Ref.~\cite{Heymans:2020gsg}).}
\label{whisker}
\end{wrapfigure}

The mismatch between the high $S_8$ value estimated by Planck assuming $\Lambda$CDM (grey contour in Fig.~\ref{2D}), $S_8=0.834\pm0.016$\footnote{All the bounds are reported at 68\% confidence level in the text.}, and the lower value preferred by cosmic shear measurements, it is known as the $S_8$ tension. This tension is above the $2\sigma$ level with KiDS-450~\cite{Kuijken:2015vca,Hildebrandt:2016iqg,Conti:2016gav,Joudaki:2016kym} ($S_8=0.745\pm0.039$) and KiDS-450+2dFLenS~\cite{Joudaki:2017zdt} ($S_8=0.742\pm0.035$), with KiDS+VIKING-450 (KV450)~\cite{Hildebrandt:2018yau} ($S_8=0.737^{+0.040}_{-0.036}$), with DES~\cite{Abbott:2017wau,Troxel:2017xyo} ($S_8=0.783^{+0.021}_{-0.025}$), and with CFHTLenS~\cite{Heymans:2012gg, Erben:2012zw,Joudaki:2016mvz}. Recently, KiDS-1000~\cite{Heymans:2020gsg} reported a $\sim3\sigma$ tension ($S_8=0.766^{+0.020}_{-0.014}$, red contour in Fig.~\ref{2D}) with Planck. This is already obvious from cosmic shear alone~\cite{Asgari:2020wuj}, but when combined with galaxy clustering, the degeneracy breaking between $\sigma_8$ and $\Omega_m$ does not change the tension level. Therefore, the combined analysis helps in pointing out that the tension, at $3.1\sigma$ in this case, is driven by $\sigma_8$ rather than $\Omega_m$. In addition, there is the Lyman-$\alpha$ result~\cite{Palanque-Delabrouille:2019iyz}, a late time probe probing scales similar to weak lensing, completely in agreement with a lower $S_8$ value and in tension at $\sim2.6\sigma$ with Planck. The tension becomes $3.2\sigma$ if we consider the combination of KV450 and DES-Y1~\cite{Joudaki:2019pmv,Asgari:2019fkq} or $3.4\sigma$ for BOSS+KV450~\cite{Troster:2019ean} ($S_8=0.728\pm0.026$, blue contour in Fig.~\ref{2D}). Preferring a higher value for the $S_8$ parameter there is also the measurement from the first-year data of HSC SSP~\cite{Hamana:2019etx}, for which $S_8=0.804^{+0.032}_{-0.029}$ (see Fig.~\ref{whisker}), but also KiDS-450+GAMA~\cite{vanUitert:2017ieu} finding $S_8=0.800^{+0.029}_{-0.027}$. Finally, in agreement with a lower value $S_8=0.703\pm0.045$ there is an estimate from the BOSS Galaxy Power Spectrum~\cite{Ivanov:2019pdj}. 

It has been pointed out in~\cite{DiValentino:2018gcu} that this tension could be related to the excess of lensing measured by Planck, mimicking a larger $S_8$. However, also ACT+WMAP~\cite{Aiola:2020azj} find a large $S_8=0.840\pm0.030$ even if it does not see a peculiar value for the lensing amplitude, while SPTpol~\cite{Henning:2017nuy} and the Planck CMB lensing ~\cite{Ade:2015zua} measurements prefer a lower value. Another possibility is the misuse of the units $h^{-1} {\rm Mpc}$ in observational cosmology in~\cite{Sanchez:2020vvb}. It might be worth mentioning that, while weak lensing analyses are carried out with a blinding procedure for KiDS, DES and HSC, the CMB analyses are either not blind or only partially blind.

\noindent {\bf Conjoined history problem --} 
The $H_0$ disagreement is correlated to the $\sigma_8$ problem, indeed the solutions proposed to alleviate the first one, are exacerbating the CMB tension with the lower $\sigma_8$ values obtained from more direct measurements, such as 
galaxy clusters using the Sunyaev-Zel'dovich effect~\cite{Ade:2015gva,Ade:2015fva,deHaan:2016qvy}, i.e. measuring the number of clusters of a certain mass M over a range of redshift.

\noindent 
For example, late time transitions preferring a higher $H_0$ value, if they match the CMB data, prefer a lower $\Omega_m$ as well, to preserve the well measured value of $\Omega_m h^2$, known as geometric degeneracy. This effect produces a modification of distances to sources, the growth of structure, and of the sound horizon and CMB anisotropies~\cite{Arendse:2019hev}, and usually results in higher $\sigma_8$ than for $\Lambda$CDM because of an extended era of matter domination. However, also early-time dark energy solutions of the $H_0$ tension increase $\sigma_8$ because they need a higher primordial curvature perturbation amplitude to offset the damping effect of the unclustered component.
Therefore, because of the mutual effects and correlations, it is important to perform a conjoined analysis, fitting with a single model a full array of data~\cite{Hill:2020osr,Benevento:2020fev,Knox:2019rjx,Evslin:2017qdn}, and not just one parameter alone. At the same time, if a model solves the $S_8$ tension (the $z=0$ value), the growth
history at different redshift, by plotting $f\sigma_8(z)$ 
directly against $H(z)$, should be checked~\cite{Linder:2016xer, DiValentino:2020kha}, because conjoint history can deviate significantly at intermediate scales.
Hence, any solution to the $S_8$ tension should pass other cosmological tests, i.e.~it should simultaneously fit the expansion and growth histories probed by Baryon Acoustic Oscillations (BAO), RSD-lensing cross correlations, galaxy power spectrum shape and void measurements~\cite{Hamaus:2020cbu}.

\noindent {\bf Solutions --} 
There are many papers investigating this tension~\cite{Troxel:2018qll,DiValentino:2018gcu,DiValentino:2015ola,DiValentino:2016hlg,Anand:2017wsj,DiValentino:2017zyq,DiValentino:2019dzu,DiValentino:2019ffd,DiValentino:2019jae,Gomez-Valent:2018nib,Gomez-Valent:2017idt,Lambiase:2018ows,Camera:2017tws,DiValentino:2016ucb,Burenin:2018nuf,Davari:2019tni,Ade:2015rim,DiValentino:2015bja,Lin:2017txz,Battye:2013xqa,Boehringer:2016bzy,Meerburg:2014bpa,Ivanov:2020ril,Klypin:2020tud,DiValentino:2017oaw,Buen-Abad:2015ova,Wang:2020dsc,Chudaykin:2017ptd,Abellan:2020pmw,Heimersheim:2020aoc,Jedamzik:2020krr,Dutta:2019pio,Sanchez:2020vvb}, but the solutions proposed are not enough to put in agreement all the cosmological available data~\cite{DiValentino:2020vhf,DiValentino:2020zio,DiValentino:2020srs}. We can distinguish the following categories:

\begin{itemize}[noitemsep,topsep=0pt]

\item  Axion monodromy inflation~\cite{Meerburg:2014bpa}.

\item Extended parameter spaces~\cite{DiValentino:2018gcu,DiValentino:2015ola,DiValentino:2016hlg,DiValentino:2017zyq,DiValentino:2019dzu} with $A_{L}>1$~\cite{Calabrese:2008rt}, i.e. using the phenomenological lensing parameter as a consistency check and determining whether it is different from unity~\cite{Aghanim:2018eyx}.

\item Active and Sterile Neutrinos~\cite{Battye:2013xqa,Boehringer:2016bzy}.

\item Interacting dark energy models, where the energy flows from the dark matter to the dark energy~\cite{DiValentino:2019ffd,DiValentino:2019jae}.

\item Decaying dark matter~\cite{Berezhiani:2015yta,Anchordoqui:2015lqa,Chudaykin:2017ptd,Abellan:2020pmw}, or Cannibal dark matter~\cite{Heimersheim:2020aoc}.

\item Minimally and non-minimally coupled scalar field models as possible alternatives for dark energy~\cite{Davari:2019tni}.

\item Modified Gravity models~\cite{Ade:2015rim,DiValentino:2015bja,Sola:2019jek,Sola:2020lba}.

\item Running vacuum models in which $\Lambda=\Lambda(H)$ is an affine power-law function of the Hubble rate~\cite{Sola:2015wwa,Gomez-Valent:2017idt,Sola:2017lxc,Gomez-Valent:2018nib,Sola:2016jky,Sola:2017jbl,Sola:2016ecz,Moreno-Pulido:2020anb}.

\item Quartessence, a single dark component mimicking both dark matter and dark energy~\cite{Camera:2017tws}.

\end{itemize}

\noindent {\bf Future --} 
In the near future, we expect percent measurements of the expansion and growth history over a large range of experiments, i.e. using maps of the Universe obtained by the Euclid satellite, measuring the peculiar motions of galaxies using Type Ia supernovae from LSST~\cite{Howlett:2017asw,Scolnic:2019apa}, considering RSD with DESI and 4MOST, or using voids~\cite{Hamaus:2020cbu}. An important role will be played by the SKA telescopes performing BAO surveys and measuring weak gravitational lensing using 21~cm intensity mapping~\cite{Pourtsidou:2014pra,Santos:2015gra,Bull:2015nra}. Additional upcoming 21~cm neutral hydrogen experiments measuring the expansion history will be CHIME and HIRAX. Finally, line-intensity mapping of emission from star-forming galaxies can be used to measure the BAO reaching percent-level constraints~\cite{Karkare:2018sar,Bernal:2019gfq} with the SPHEREx satellite or the ground-based COMAP instrument. All of these efforts will either reveal a systematic cause or harden the tension to strong statistical significance informing the theories mentioned above and guiding any extension/overhaul of the standard model.

\clearpage

\bibliographystyle{utphys}
\bibliography{H0}

\end{document}